\documentclass[aps,prl,amsmath,amssymb,twocolumn,preprintnumbers,superscriptaddress,showpacs,lengthcheck]{revtex4-1}

\newcommand{\ncco}{$\textrm{Nd}_{2-x}\textrm{Ce}_x\textrm{Cu}\textrm{O}_{4}$}

\usepackage{graphicx}
\usepackage{dcolumn}
\usepackage{bm}
\usepackage{relsize}
\usepackage{natbib}
\usepackage{float}
\begin{document}

\title{Charge ordering in the electron-doped superconductor \ncco}

\author{Eduardo H. da Silva Neto\textsuperscript{\textdaggerdbl}}
\email[]{ehda@phas.ubc.ca}
\affiliation{\mbox{Department of Physics \& Astronomy, University of British Columbia, Vancouver, British Columbia, V6T 1Z1, Canada}}
\affiliation{\mbox{Quantum Matter Institute, University of British Columbia, Vancouver, British Columbia, V6T 1Z4, Canada}}
\affiliation{Max Planck Institute for Solid State Research, Heisenbergstrasse 1, D-70569 Stuttgart, Germany}

\author{Riccardo Comin\textsuperscript{\textdaggerdbl}}
\affiliation{\mbox{Department of Physics \& Astronomy, University of British Columbia, Vancouver, British Columbia, V6T 1Z1, Canada}}
\affiliation{\mbox{Quantum Matter Institute, University of British Columbia, Vancouver, British Columbia, V6T 1Z4, Canada}}

\author{Feizhou He}
\affiliation{Canadian Light Source, Saskatoon, Saskatchewan S7N 2V3, Canada}

\author{\mbox{Ronny Sutarto}}
\affiliation{Canadian Light Source, Saskatoon, Saskatchewan S7N 2V3, Canada}

\author{Yeping Jiang}
\affiliation{CNAM and Department of Physics, University of Maryland, College Park, Maryland 20742, USA}

\author{Richard L. Greene}
\affiliation{CNAM and Department of Physics, University of Maryland, College Park, Maryland 20742, USA}

\author{George A. Sawatzky}
\affiliation{\mbox{Department of Physics \& Astronomy, University of British Columbia, Vancouver, British Columbia, V6T 1Z1, Canada}}
\affiliation{\mbox{Quantum Matter Institute, University of British Columbia, Vancouver, British Columbia, V6T 1Z4, Canada}}

\author{Andrea Damascelli}
\email[]{damascelli@physics.ubc.ca}
\affiliation{\mbox{Department of Physics \& Astronomy, University of British Columbia, Vancouver, British Columbia, V6T 1Z1, Canada}}
\affiliation{\mbox{Quantum Matter Institute, University of British Columbia, Vancouver, British Columbia, V6T 1Z4, Canada}}


\begin{abstract}
In cuprate high-temperature superconductors, an antiferromagnetic Mott insulating state can be destabilized toward unconventional superconductivity by \emph{either} hole- or electron-doping. In addition to these two electronic phases there is now a copious amount of evidence that supports the presence of a charge ordering (CO) instability competing with superconductivity inside the pseudogap state of the hole-doped (\emph{p}-type) cuprates, but so far there has been no evidence of a similar CO in their electron-doped (\emph{n}-type) counterparts. Here we report resonant x-ray scattering (RXS) measurements which demonstrate the presence of charge ordering in the \emph{n}-type cuprate \ncco~near optimal doping. Remarkably we find that the CO in \ncco~occurs with similar periodicity, and along the same direction, as the CO in \emph{p}-type cuprates. However, in contrast to the latter, the CO onset in \ncco~is higher than the pseudogap temperature, and is actually in the same temperature range where antiferromagnetic fluctuations are first detected -- thereby showing that CO and antiferromagnetic fluctuations are likely coupled in \emph{n}-type cuprates. Overall our discovery uncovers a missing piece of the cuprate phase diagram and opens a parallel path to the study of CO and its relationship to other phenomena, such as antiferromagnetism (AF) and high-temperature superconductivity.
\end{abstract}

\maketitle

\begin{figure*}
\includegraphics[width=175mm]{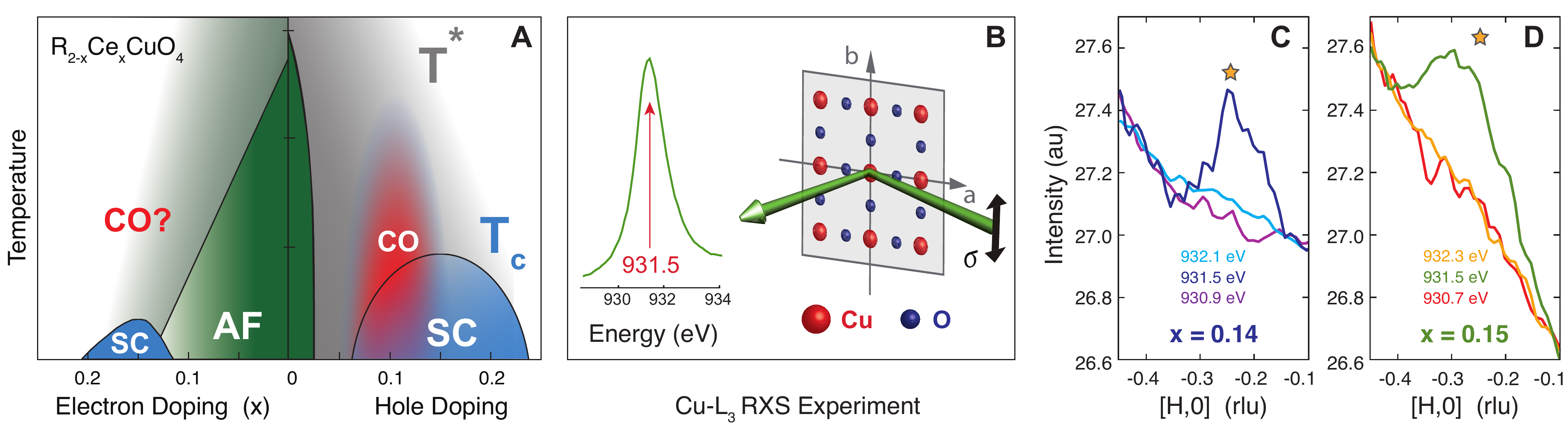}
\caption{\label{fig:1}\textbf{Charge ordering in electron doped cuprates.} (\textbf{A}) Temperature-doping phase diagram for the cuprates, including the AF parent state (green), the superconductivity (SC, blue, $T_c$), and distinct $n$-type (faded green) and $p$-type (gray) pseudogap phases. In $p$-type cuprates CO is observed, though its $n$-type counterpart has not yet been observed. (\textbf{B}) The Cu $L_3$ absoprtion edge at $931.5$\,eV ($2p \rightarrow 3d$ transition) and a schematic of the scattering geometry (right, see text for details). (\textbf{C-D}) On- and off-resonance $\theta$-scans at $22$\,K showing the RXS diffraction signal as a function of in-plane momentum transfer (H) along the Cu-O bond direction (see B). To provide a better comparison, the off-resonance scans were rescaled to match the tails of the on-resonance $\theta$-scans. A similar resonant peak is also observed along the K direction (not shown). The yellow stars mark the H-values of highest intensity for the two samples (obtained from Fig.\,\ref{fig:3}).}
\end{figure*}

Given the abundance of possible electronic phases in the cuprates, the immediate problem becomes to determine the relevance of these phenomena to the superconducting pairing mechanism. One path toward solving this problem is to determine which phenomena are universal to both $p$- and $n$-type cuprates and how they might depend on the specifics of the participating hole- or electron-like states \cite{armitage_progress_2010}. Notably, CO has recently taken the forefront of cuprate research, but it has so far not been detected in $n$-type cuprates (Fig.\,\ref{fig:1}A). Early evidence for a CO in the cuprates came from the detection in La-based cuprates of a periodic organization of spins and charge known as stripes  \cite{tranquada_evidence_1995, Abbamonte_2005, Fink_2009, Fink_2011}, where the charge is periodic every four lattice constants along the Cu-O bond direction. More recently, following a growing amount of evidence for Fermi surface reconstruction from quantum oscillations \cite{doiron-leyraud_quantum_2007, sebastian_2011}, nuclear magnetic resonance \cite{Wu_2011} and x-ray scattering measurements \cite{Ghiringhelli_CDW_2012, Chang_CDW_2012} have directly shown the presence of a similar CO competing with superconductivity in Y-based cuprates. The opportunity to directly probe CO in reciprocal space has further propelled several resonant x-ray scattering measurements of the Y-based family \cite{Achkar_2012, Santi_2013, Santi_2014}, as well as the detection of CO in Bi-cuprates \cite{Comin_CDW_2014, daSilvaNeto_CDW_2014, Hashimoto_2014} -- substantiating earlier surface evidence by scanning tunneling microscopy \cite{Hoffman_2002,Vershinin_2004, Howald_2003, Wise_2008} -- and also in the single-layer Hg-compound \cite{Tabis_2014}.



In particular, studies of Bi-based cuprates, for which a considerable amount of angle-resolved photoemission spectroscopy (ARPES) data is available, show that the CO wavevector connects the ends of the Fermi arcs \cite{Comin_CDW_2014, daSilvaNeto_CDW_2014} -- an observation that links the existence of CO to the pseudogap in hole-doped systems. Additionally, doping dependent measurements on bi-layer systems \cite{parker_fluctuating_2010, Ghiringhelli_CDW_2012,  Santi_2014, Huecker_2014} find charge ordering to be most pronounced in a region of hole-doping near $x=1/8$, where stripes are predominant in La-based cuprates \cite{tranquada_evidence_1995, Abbamonte_2005}. These results raise questions of whether the particular phenomenology of the hole-doped cuprates such as the pseudogap-induced Fermi arcs, or the propensity toward stripe formation, are necessary ingredients for CO formation, or whether CO is a generic electronic property of the CuO$_2$ layer which is ubiquitous to all cuprates including $n$-type materials.

Here we report resonant x-ray scattering (RXS) measurements on the electron-doped cuprate superconductor \ncco~\cite{SM2}. Our studies were performed on samples with doping levels ($x=0.14\pm0.01$ and $x=0.15\pm0.01$), for which quantum oscillations indicate a small Fermi surface \cite{Helm_evolution_2009, Helm_magnetic_2010}. We also use the standard scattering geometry shown in Fig.\,\ref{fig:1}B \cite{SM2}, similar to previous RXS measurements on the cuprates \cite{Ghiringhelli_CDW_2012, Comin_CDW_2014, daSilvaNeto_CDW_2014}. The tetragonal b-axis of the sample is positioned perpendicular to the scattering plane allowing the in-plane components of momentum transfer to be accessed by rotating the sample around the b-axis ($\theta$-scan). For RXS measurements the energy of the incoming photons is fixed to the maximum of the Cu-$L_3$ absorption edge, which is at $E\simeq931.5$\,eV (Fig.\,\ref{fig:1}B). 

Our main finding is summarized in Figs.\,\ref{fig:1}C-D which display a RXS peak at an in-plane momentum transfer of $\text{H} \simeq -0.24$ rlu (reciprocal lattice units) along the Cu-O bond direction -- remarkably similar in periodicity and direction to the RXS peaks found in the hole-doped materials \cite{Abbamonte_2005, Fink_2009, Fink_2011, Ghiringhelli_CDW_2012, Chang_CDW_2012, Achkar_2012, Santi_2013, Huecker_2014, Santi_2014, Comin_CDW_2014, daSilvaNeto_CDW_2014, Hashimoto_2014, Tabis_2014}. The use of photons tuned to the Cu-$L_3$ edge is expected to greatly enhance the sensitivity in our measurement to charge modulations involving the valence electrons in the CuO$_2$ planes \cite{Abbamonte_2005}. In particular Figs.\,\ref{fig:1}C-D show that, as the photon energy is tuned away from resonance, the distinct peak near $\text{H}=-0.24$ disappears, thus confirming its electronic origin \cite{SM2}. This shows for the first time the presence of charge ordering in an electron-doped cuprate.

\begin{figure}[b]
\includegraphics[width=85mm]{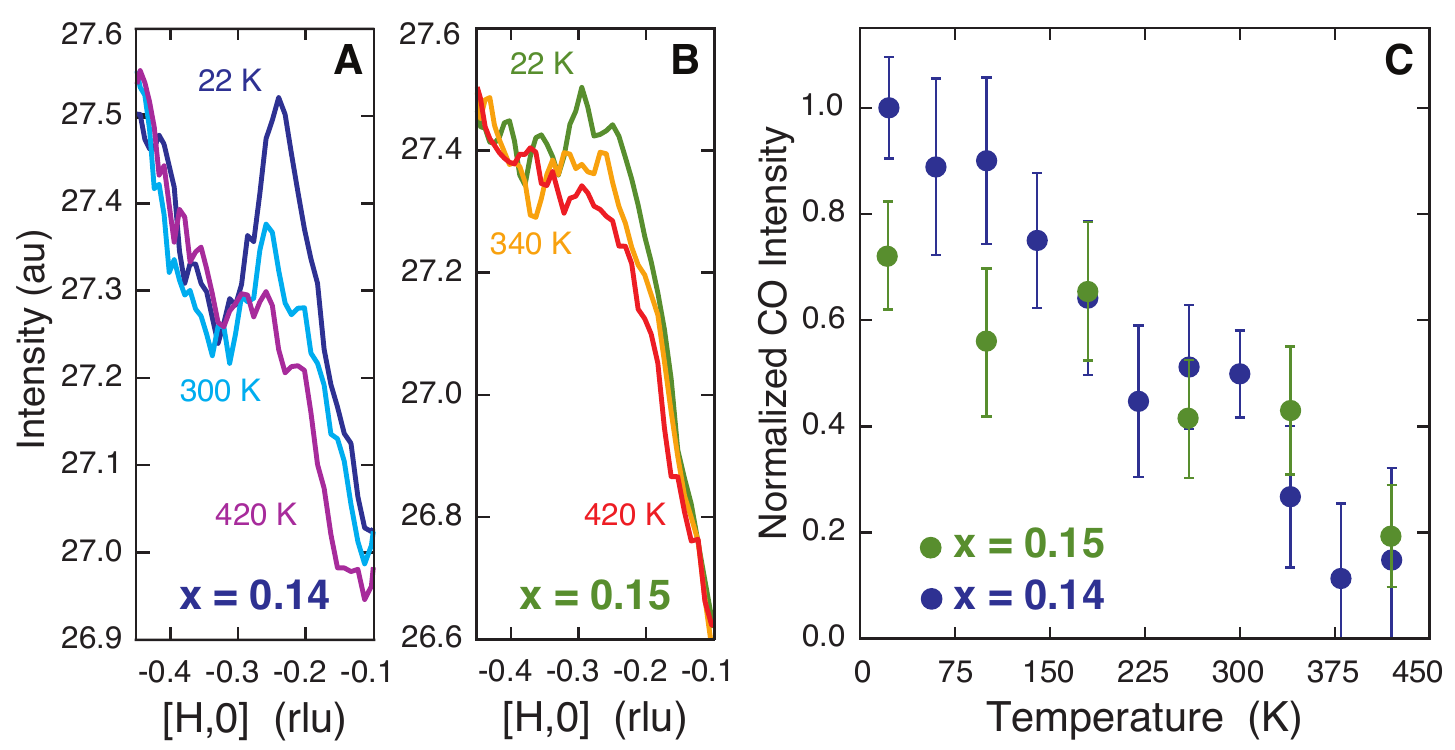}
\caption{\label{fig:2} \textbf{Temperature dependence of the CO.} (\textbf{A}-\textbf{B}) On-resonance $\theta$-scans for $x=0.14$ and $x=0.15$ samples at select temperatures show that the onset of the charge ordering occurs above $300$\,K. (\textbf{C}) Temperature dependence of the RXS intensity for the two samples in A and B obtained from the maxima of the background-subtracted peaks. The intensity in (C) is normalized to maximum value between the two samples and the error bars represent the standard errors from Lorentzian fits to the background-subtracted peaks \cite{SM2}.}
\end{figure}

Further insights into charge ordering formation are obtained by temperature-dependent measurements. Figures \ref{fig:2}A-B show that the distinct CO peak observed at low temperatures weakens as the temperature is raised but only disappears above $300$\,K. Though a temperature evolution is clearly seen in the the raw data (Figs.\,\ref{fig:2}A-B and S3), the small size of the peak relative to the high-temperature background precludes a precise determination of an onset temperature. Nevertheless, within the detection limits of the experiment, the CO seems to gradually develop with lowering of temperature starting around $340$\,K (see Fig.\,\ref{fig:2}C). It is important to note that this temperature is much higher than the pseudogap onset in \ncco~($\sim$$80\text{-}170$\,K in the $x=0.14$-$0.15$ doping range \cite{armitage_progress_2010, Onose_2004, Motoyama_Spin_2007}), in clear contrast to observations in hole-doped cuprates where the $p$-type pseudogap either precedes or matches the emergence of CO \cite{parker_fluctuating_2010, Ghiringhelli_CDW_2012, Chang_CDW_2012,Achkar_2012, Santi_2013, Huecker_2014, Comin_CDW_2014, daSilvaNeto_CDW_2014, Tabis_2014, Santi_2014}. But of course this dichotomy is not completely unexpected given that the pseudogaps observed in $p$- and $n$-type cuprates are dissimilar in many ways \cite{armitage_progress_2010}. In particular, the $n$-type pseudogap has been associated with the buildup of AF correlations that first appear below $320$\,K (for $x=0.145$ samples), as determined by inelastic neutron scattering measurements \cite{Kyung_2004, Onose_2004, Motoyama_Spin_2007}. Interestingly, the temperature evolution of the CO resembles the soft onset of AF correlations \cite{Motoyama_Spin_2007} -- an observation that suggests a close connection between CO and AF fluctuations in electron-doped cuprates.

\begin{figure*}[t]
\centering{
\includegraphics[width=185mm]{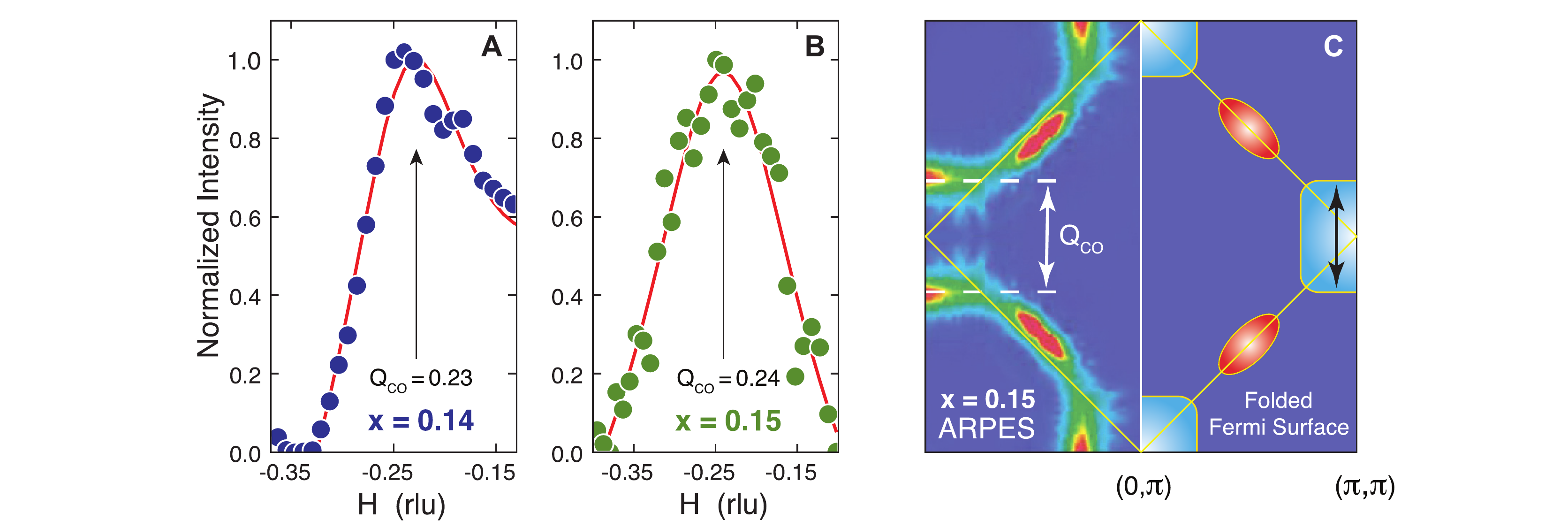}
\caption{\label{fig:3}\textbf{Electronic origin of the CO.} (\textbf{A}-\textbf{B}) CO peak extracted by subtraction of the highest-temperature $\theta$-scan from an average of the lowest temperatures ($22$ to $180$\,K). A fit of the data to a Lorentzian plus linear background function (red line) is used to indicate the H value of highest intensity, which is $-0.23\pm0.04$ ($-0.24\pm0.04$) rlu for the $x=0.14$~($x=0.15$) sample. The extracted peaks in A and B are normalized to their respective maxima. (\textbf{C}) Fermi surface of \ncco~($x=0.15$) measured by ARPES \cite{Armitage_anomalous_2001} (left) and a schematic of the expected Fermi surface reconstruction (right) due to AF folding (yellow diamond). The folded Fermi surface is composed of hole (red) and electron (cyan) pockets. The arrows (white and black) and dashed lines represent $Q_{\text{CO}}=0.24$\,rlu, and connect either the parallel segments of the Fermi surface near $(\pi,0)$, or the intersection with the AF zone boundary.}
}
\end{figure*}

We now use the available knowledge of the Fermi surface of \ncco~to further investigate the connection between AF and CO formation. Though broad, we find that the CO peak is centered around $Q_{\text{CO}}=0.23\pm0.04$ and $Q_{\text{CO}}=0.24\pm0.04$ for $x=0.14$ and $x=0.15$, respectively (Fig.\,\ref{fig:3}A-B). Comparison of $Q_{\text{CO}}$ to the Fermi surface topology measured by ARPES (Fig.\,\ref{fig:3}C, left half) shows that its value is consistent with scattering between the parallel segments near $(\pi,0)$. Due to the relative robustness of the AF phase in $n$-type cuprates, the Fermi surface has often been interpreted to undergo $(\pi,\pi)$ folding along the AF zone boundary -- a scenario which is consistent with both ARPES \cite{Armitage_anomalous_2001} and quantum oscillation results \cite{Helm_evolution_2009}. In this context $Q_{\text{CO}}$ would connect opposite sides of electron pockets centered at $(\pi,0)$ (Fig.\,\ref{fig:3}C, right half). Alternatively, $Q_{\text{CO}}$ might instead connect the intersections between the AF zone boundary and the underlying Fermi surface, the so-called hot spots where the effect of AF scattering and the pseudogap are maximal \cite{Armitage_anomalous_2001}. However, the conventional expectation that the onset of CO above room temperature should gap the Fermi surface seems to contradict both scenarios since the pseudogap only opens at the hot spots below $180$\,K, whereas no gapping is observed near $(\pi, 0)$ above the superconducting transition \cite{armitage_progress_2010} - suggesting that Fermi surface nesting might not be the origin of the CO. Unfortunately, however, this kind of comparison between temperature scales might be rendered inconclusive by the possibility that the CO never becomes sufficiently long-ranged, or large enough in amplitude, to induce a detectable reconstruction of the Fermi surface (at least in the absence of an applied magnetic field). Indeed, the widths of the CO peaks shown in \mbox{Fig.\,\ref{fig:3}A-B} indicate a short correlation length ($25$ to $35$\,\AA), again similar to what has been observed in Bi-based cuprates \cite{Comin_CDW_2014, daSilvaNeto_CDW_2014, Hashimoto_2014}. Perhaps further measurements, spanning larger doping ranges, will be able to test exactly which momentum states are involved in the CO, though the broadness of the CO peak in reciprocal space might ultimately limit the precision to which the location of $Q_{\text{CO}}$ on the Fermi surface can be determined. 



The fact that CO never develops into a long-ranged electronic ground state might also hinder the ability of transport or thermodynamic probes to detect it. However, we find that the presence of CO might be relevant to the interpretation of experiments that probe the inelastic excitations of \ncco. We start by observing that the value of $Q_{\text{CO}}$ is consistent with the phonon anomaly near $\text{H} \simeq 0.2$ observed by inelastic x-ray scattering in \ncco~\cite{dAstuto_2002}. More recently, Hinton \emph{et al.}\,\cite{Hinton_TRR_2013} reported time-resolved reflectivity (TRR) studies which show the presence of a fluctuating order competing with superconductivity, though they could not determine which electronic degrees of freedom (i.e. charge or spin) were responsible for such order. Additionally, resonant inelastic x-ray scattering (RIXS) measurements by Lee \emph{et al.}\,\cite{Lee_asymmetry_2013} and Ishii \emph{et al.}\,\cite{Ishii_Soin_2014} have recently showed the presence of an inelastic mode -- present above a minimum energy transfer of $300\pm30$\,meV (comparable to the pseudogap \cite{Armitage_anomalous_2001}) -- which is distinct from the well-characterized AF fluctuations reminiscent of the Mott-insulating parent state \cite{Yamada_Spin_2003, Motoyama_Spin_2007, Lee_asymmetry_2013, Ishii_Soin_2014}. While the authors of the later study ascribed this new mode to particle-hole charge excitations, Lee \emph{et al.} proposed that this new mode might be the consequence of a then unknown broken symmetry -- a scenario supported by their observation that this mode disappears above $270$\,K for $x=0.166$. Here, our discovery of charge ordering in \ncco~might provide the missing piece of information to properly interpret these aforementioned studies by identifying the actual broken symmetry.  


Finally, on a fundamental level, some degree of electron-hole asymmetry should be expected in the cuprate phase diagram. In fact, whereas doped hole states below the charge transfer gap have a strong O-$2p$ character, $n$-type doping creates low-energy electronic states of predominantly Cu-$3d$ character in the upper Hubbard band \cite{ZSA_1985,Chen_XAS_1991,Pellegrin_XAS_1993}. This dichotomy, together with recent RXS reports of a bond-centered CO in $p$-type materials \cite{Comin_2014}, suggests that the $n$-type CO observed here may instead be centered on the Cu sites -- an idea that requires further investigation. However, despite this underlying electron-hole asymmetry, the CO uncovered in \ncco~by our study shows several similarities to its $p$-type equivalent, such as its direction, periodicity, and short correlation length \cite{Comin_CDW_2014, daSilvaNeto_CDW_2014}. In addition, our observation of a connection between the onset of CO and AF fluctuations suggests that the latter might generally lead to an accompanying intertwined charge order in unconventional superconductors, regardless of which orbitals are involved in the CO \cite{Fradkin_intertwined, Davis_DHLee_2013}. If such is the case, detailed studies will be necessary to understand the role of antiferromagnetism in charge order formation, perhaps even beyond the cuprates. Nevertheless, our discovery of charge ordering in $n$-type cuprates not only expands the universality of this phenomenon to the electron-doped side of the phase diagram, but it also provides a new avenue to understand its microscopic origin by exploiting the differences between $p$- and $n$-type cuprates. 

\vspace{3 mm}
\centerline{\textbf{Acknowledgments}}
\vspace{1 mm}
We thank Jongmoon Shin for the WDX measurements and Shanta Saha for the susceptibility measurements. We also acknowledge N.\,P. Armitage, S.\,A. Kivelson, A.-M.-S. Tremblay for fruitful discussions. This work was supported by the Max Planck - UBC Centre for Quantum Materials, the Killam, Alfred P. Sloan, Alexander von Humboldt, and NSERC's Steacie Memorial Fellowships (A.D.), the Canada Research Chairs Program (A.D., G.A.S.), NSERC, CFI, and CIFAR Quantum Materials. Work at UMD was supported by the NSF DMR 1104256. E.H.d.S.N. acknowledges support from the CIFAR Global Academy. R.C. acknowledges the support from the CLS Graduate Student Travel Support Program. All of the x-ray experiments were performed at beamline REIXS of the Canadian Light Source \cite{Hawthorn_CLS}, which is funded by the CFI, NSERC, NRC, CIHR, the Government of Saskatchewan, WD Canada, and the University of Saskatchewan.

\vspace{5 mm}
\textsuperscript{\textdaggerdbl} \small{These authors contributed equally to this work.}


\bibliographystyle{apsrev4-1}
\bibliography{bib_lib}
\end{document}